\documentclass[journal,twoside,web]{ieeecolor2}
\usepackage{generic}
\usepackage{amsmath,amssymb,amsfonts,flafter,multirow,tabularx,booktabs}
\usepackage{algorithmic}
\usepackage{graphicx}
\usepackage{textcomp}
\usepackage{amsmath}
\usepackage{hyperref}
\usepackage{xcolor}

\usepackage{hyperref}

\def\BibTeX{{\rm B\kern-.05em{\sc i\kern-.025em b}\kern-.08em
    T\kern-.1667em\lower.7ex\hbox{E}\kern-.125emX}}
\begin{document}
\title{TC-KANRecon: High-Quality and Accelerated MRI Reconstruction via Adaptive KAN Mechanisms and Intelligent Feature Scaling}
\author{Ruiquan Ge, Xiao Yu, Yifei Chen, Guanyu Zhou, Fan Jia, Shenghao Zhu, Junhao Jia\\  Chenyan Zhang, Yifei Sun, Dong Zeng, Changmiao Wang, Qiegen Liu, Shanzhou Niu
\thanks{This work was supported by the Open Project Program of the State Key Laboratory of CAD\&CG, Zhejiang University (Grant No.A2410), National Natural Science Foundation of China (No. 61702146, 62076084, U22A2033, U20A20386), Zhejiang Provincial Natural Science Foundation of China (No.LY21F020017, 2022C03043), Guangdong Basic and Applied Basic Research Foundation  (No. 2022A1515110570), Guangxi Key R\&D Project (No. AB24010167), Innovation Teams of Youth Innovation in Science, Technology of High Education Institutions of Shandong Province (No.2021KJ088), Shenzhen Science and Technology Program (No. KCXFZ20201221173008022), and Shenzhen Stability Science Program 2022 (2023SC0073). (Corresponding author: Dong Zeng, Changmiao Wang.)}
\thanks{R. Ge, X. Yu, Y. Chen, G. Zhou, F. Jia, S. Zhu, J. Jia, C. Zhang and Y. Sun are with Hangzhou Dianzi University, Hangzhou, 310018, China (e-mail:(gespring, 22320313, chenyifei, 23320307, 18120401, 22320220, 21321108, 21321231, 22320232)@hdu.edu.cn).}
\thanks{D. Zeng is with Southern Medical University, Guangdong, 510000, China (e-mail: zd1989@smu.edu.cn).}
\thanks{C. Wang is with Shenzhen Research Institute of Big Data, Shenzhen, 518172, China (e-mail: cmwangalbert@gmail.com).}
\thanks{Q. Liu  is with Department of Information Engineering, Nanchang University, Nanchang 330031, China (e-mail:liuqiegen@ncu.edu.cn).}
\thanks{S. Niu is with School of Mathematics and Computer Science, Gannan Normal University, Ganzhou 341000, China (e-mail: szniu@gnnu.edu.cn).}}

\maketitle

\begin{abstract}
Magnetic Resonance Imaging (MRI) has become essential in clinical diagnosis due to its high resolution and multiple contrast mechanisms. However, the relatively long acquisition time limits its broader application. To address this issue, this study presents an innovative conditional guided diffusion model, named as TC-KANRecon, which incorporates the Multi-Free U-KAN (MF-UKAN) module and a dynamic clipping strategy. TC-KANRecon model aims to accelerate the MRI reconstruction process through deep learning methods while maintaining the quality of the reconstructed images. The MF-UKAN module can effectively balance the tradeoff between image denoising and structure preservation. Specifically, it presents the multi-head attention mechanisms and scalar modulation factors, which significantly enhances the model's robustness and structure preservation capabilities in complex noise environments. Moreover, the dynamic clipping strategy in TC-KANRecon adjusts the cropping interval according to the sampling steps, thereby mitigating image detail loss typicalching the visual features of the images. Furthermore, the MC-Model incorporates full-sampling k-space information, realizing efficient fusion of conditional information, enhancing the model's ability to process complex data, and improving the realism and detail richness of reconstructed images. Experimental results demonstrate that the proposed method outperforms other MRI reconstruction methods in both qualitative and quantitative evaluations. Notably, TC-KANRecon method exhibits excellent reconstruction results when processing high-noise, low-sampling-rate MRI data. Our source code is available at \href{https://github.com/lcbkmm/TC-KANRecon}{https://github.com/lcbkmm/TC-KANRecon}.
\end{abstract}


\section{Introduction}
\IEEEPARstart{M}agnetic Resonance Imaging (MRI) is a crucial medical imaging technology for clinical diagnosis and research due to its high resolution and multiple contrast mechanisms. MRI provides precise anatomical and functional information, making it essential for diagnosing neurological diseases, cardiovascular conditions, and cancer. Compared with other imaging modalities, MRI is radiation-free and offers high soft tissue contrast, leading to its widespread use in neuroimaging, cardiac imaging, and oncology~\cite{a1otazo2021mr}. Despite its advantages, MRI faces significant challenges, particularly its long acquisition time. Time-consuming scan may increase patient discomfort and limit the efficiency of equipment usage. Prolonged scan times often result in motion artifacts that degrade image quality, especially in elderly and pediatric patients who struggle to remain still for extended periods. Additionally, longer scan times escalate medical costs and reduce the equipment's turnover efficiency~\cite{a2lustig2008compressed}. To address these issues, researchers have developed various methods to accelerate MRI acquisition and reconstruction. The goal is to shorten scan times while maintaining high image quality, thereby improving patient comfort and optimizing the use of MRI technology.

The traditional MRI acquisition process is slow primarily due to the need to comprehensively sample k-space data, which represents the frequency domain of MR images. The final image is obtained by performing an inverse Fourier transform on k-space data. To acquire high-quality images, comprehensive sampling of the entire k-space is typically required, thus extending scan times. This prolonged acquisition period places a burden on patients and increases the likelihood of motion artifacts, which can reduce diagnostic accuracy. To mitigate this issue, undersampling k-space data has become a common technique to accelerate MRI acquisition. This method involves undersampling the k-space signal and using reconstruction algorithms to recover the signal~\cite{a3hollingsworth2015reducing}. Generally, the problem is formulated as follows:
\begin{equation}
\label{eq1}
 \begin{aligned}
    &\boldsymbol{y}_{recon} = \arg\underset{{y}}{\min} \parallel \boldsymbol{\mathcal{M}\boldsymbol{\mathcal{F}y}} - \boldsymbol{x}_{obs} \parallel + \boldsymbol{\lambda} \text{\textbf{R}}(\boldsymbol{y}), \\
    &s.t. \quad \boldsymbol{x}_{obs} = \boldsymbol{\mathcal{M}x}_{full},
\end{aligned}  
\end{equation}
where $\boldsymbol{x}_{full}$ and $\boldsymbol{x}_{obs}$ denote the fully-sampled and undersampled k-space signal, respectively, $\boldsymbol{\mathcal{M}}$ represents the undersampling mask, and $\boldsymbol{\mathcal{F}}$ is the Fourier operator. The goal is to find a MR image $\boldsymbol{y}$ such that its k-space content $\boldsymbol{\mathcal{M}}\mathcal{\boldsymbol{\mathcal{F}y}}$ aligns with $\boldsymbol{x}_{obs}$, often referred to as the data consistency term. Furthermore, $\boldsymbol{y}_{recon}$ should adhere to certain prior knowledge about MR images, as expressed by the regularization term $ \textbf{R}(\cdot) $, which is subject to numerous innovations.

Although undersampling reduces the amount of acquired data and shortens scan times, it violates the Nyquist-Shannon sampling theorem, potentially introducing aliasing artifacts. To address this, researchers have proposed Compressed Sensing (CS) technology~\cite{a4haldar2010compressed} to mitigate aliasing artifacts caused by undersampling. CS formulates image reconstruction as an optimization problem, incorporating assumptions of sparsity and incoherence. However, the complexity and limitations of CS have hindered its widespread application due to the need for multiple iterative computations and substantial computational resources. Additionally, the effectiveness of CS methods heavily relies on empirical hyperparameter selection, necessitating individualized adjustments under different scanning conditions, a time-consuming and laborintensive process~\cite{chen2022ai}. To further improve MRI acceleration capabilities, researchers have explored various methods such as low-rank constraints~\cite{a5majumdar2015improving}, adaptive sparse modeling~\cite{a7patel2011gradient},  applying the TVJ1-ESPIRiT method in combination with PI and CS~\cite{10689642} and parallel imaging techniques~\cite{a8liang2009accelerating}. These methods enhance image reconstruction quality and speed through diverse technical approaches but still face challenges with aliasing artifacts at high acceleration factors~\cite{a9ravishankar2010mr}. Parallel imaging techniques, which use multiple receiver coils to simultaneously acquire signals, can enhance acquisition speed and image quality. However, their effectiveness depends on coil configuration and imaging parameter selection, requiring experienced operators.

In recent years, the application of artificial intelligence technology in the field of MRI has infused new vitality into MRI acceleration and reconstruction techniques. Through training, deep learning models are capable of reconstructing high-quality images from undersampled k-space data, drastically simplifying the complex parameter tuning processes of traditional methods. Their efficient computational capabilities further drastically reduce image reconstruction time, enhancing the efficiency of medical services. However, despite breakthroughs in existing deep learning technologies such as diffusion models, challenges persist. Noise sensitivity is particularly problematic, as device noise, patient movement, and other factors in MRI data can easily interfere with image quality, leading models to mistakenly learn noise, resulting in artifacts or blurred images that compromise diagnostic accuracy. Additionally, inadequate preservation of structural information poses a significant challenge. The fine anatomical structures in MR images are crucial for disease diagnosis, yet existing methods often struggle to retain these details while achieving rapid reconstruction. Furthermore, the lack of robustness in deep learning models limits their clinical application. The significant variations in MRI data among different patients, scanning protocols, and devices require models to possess strong generalization capabilities. However, current models are often trained on specific datasets, and their adaptability to new environments needs improvement. To address these challenges, we innovatively propose the Multi-Free U-KAN (MF-UKAN) module and its complementary dynamic cropping strategy, significantly enhancing the precision, efficiency, and flexibility of diffusion models in MR image reconstruction tasks. Our method offers several key contributions and advantages:
\begin{itemize}
    \item To balance denoising with the preservation of structural information, we introduced the MF-UKAN module. This module employs multi-head attention mechanisms and scalar modulation factors, providing fine-grained control over backbone and skip features. This design greatly enhances the model's robustness in noisy environments and also improves the retention of structural details in reconstructed images.
    \item We introduced an innovative dynamic clipping strategy to overcome the limitations of traditional cropping methods, which often restrict image diversity. Our strategy dynamically adjusts cropping interval boundaries based on sampling steps, effectively reducing image detail loss. This results in images with richer visual features and better brightness distribution.
    \item We integrated fully sampled k-space information with MRI data in the MC-Model module, processing it through the encoder stage of the MF-UKAN module. This combination enhances the model's capability to handle complex data and aids in generating targeted MR images, thus improving the realism and detail in reconstructed images.
    \item We conducted extensive comparative experiments on two public datasets. Our results indicate that our method outperforms other state-of-the-art MRI reconstruction methods, particularly in terms of performance when processing high-noise, low-sampling-rate MRI data. Ablation experiments further validated the critical role of each module in the TC-KANRecon model, confirming the effectiveness and necessity of our design.
    
\end{itemize}

\section{Related work}
With the rapid advancement of deep learning, models leveraging this technology have become increasingly prominent in MRI reconstruction, demonstrating exceptional performance. Early approaches~\cite{b3guo2020anatomic} utilized single feedforward Convolutional Neural Networks (CNNs), such as SRCNN~\cite{b4dong2014learning} and U-Net~\cite{b5ronneberger2015u}, to map undersampled k-space data to fully sampled images in an end-to-end manner. More sophisticated models have since emerged, adopting iterative architectures that break down the reconstruction process into several learnable optimization stages. For instance, Sun \textit{et al.}~\cite{b6yang2017admm} introduced ADMM-Net for MRI reconstruction, which represents each stage as an iteration of the Alternating Direction Method of Multipliers (ADMM) algorithm~\cite{b7boyd2011distributed}. Similarly, Dar \textit{et al.}~\cite{b8dar2020prior} employs conditional generative adversarial networks for the joint recovery of undersampled multi-contrast acquisitions. To emulate iterative dictionary learning methods, Schlemper \textit{et al.}~\cite{b9schlemper2017deep} and Dar \textit{et al.}~\cite{b10dar2020transfer} proposed deep cascaded CNN architectures, which perform convolution operations in image space through multiple residual learning blocks, Zeng \textit{et al.}~\cite{zeng2020comparative} applied CNN in MRI reconstruction. Qin \textit{et al.}~\cite{b11qin2018convolutional} further advanced this concept by introducing a Convolutional Recurrent Neural Network (CRNN) architecture that leverages time-series dependencies and iterative optimization benefits. Subsequent improvements to CRNN designs by Wang \textit{et al.}~\cite{b12chen2022pyramid}, Chen \textit{et al.}~\cite{b13chen2020mri}, and Guo \textit{et al.}~\cite{b14guo2021over}  incorporated recursive pyramid layers, neural ODEs, and overcomplete network architectures in the hope of further enhancing reconstruction quality through the above strategies. In addition, Machado \textit{et al.}~\cite{10269012}   proposes method adds the undersampling factor in the reconstruction process and combining it with other tasks, achieving faster and better reconstruction results.

Generative Adversarial Networks (GANs) have shown promise in this field due to their ability to learn data distributions more rapidly than traditional CNN models. Mardani \textit{et al.}~\cite{b15mardani2018deep} combined deep GAN and CS to reduce high-frequency noise and enhance zero-filled MR images. Quan \textit{et al.}~\cite{b16quan2018compressed} developed a dual-bench generator using cyclic data consistency loss and generative adversarial loss for accurate reconstruction of undersampled data. Li \textit{et al.}~\cite{b17li2019segan} proposed SEGAN, which employs patch correlation regularization to recover structural information both locally and globally. Murugesan \textit{et al.}~\cite{b18murugesan2019recon} introduced Recon-GLGAN, a framework that combines global and local context information through a generator and a context discriminator. Shaul \textit{et al.}~\cite{b19shaul2020subsampled} developed a software-based GAN framework to estimate missing k-space samples, accelerating brain MRI acquisition.

While GAN-based methods are renowned for their ability to generate realistic images, they often suffer from a lack of diverse representation, which can impede their performance in image reconstruction. Recent advancements in diffusion models offer promising solutions to this limitation. These models convert Gaussian noise into image samples through a multi-step procedure that directly maps the correlations within the data distribution. By integrating learned priors with imaging operators during the inference stage, diffusion models enable repeated projections for improved reconstruction. For example, Jalal \textit{et al.}~\cite{b20jalal2021robust} and Luo \textit{et al.}~\cite{b21luo2022mri} introduced the use of score functions and Langevin dynamics for sampling. Similarly, Song \textit{et al.}~\cite{b23song2021solving} utilized predictors to address stochastic differential equations in the methodology. Peng \textit{et al.}~\cite{b24peng2022towards} developed DiffuseRecon, a diffusion model-based MRI reconstruction approach that eliminates the need for additional training for different acceleration factors. Meanwhile, the CDPM model proposed by Cao \textit{et al.}~\cite{b25cao2022accelerating} effectively preserves the complex-valued information critical in MRI data. Gungor \textit{et al.}~\cite{b26gungor2023adaptive} created Adaptive Diffusion Priors, such as AdaDiff, to further enhance reconstruction performance during inference. Cao \textit{et al.}~\cite{b27cao2024high} also designed a high-frequency DDPM model to retain high-frequency information in MRI data.  Despite these advancements, diffusion-based methods have not completely overcome the inherent limitations of the U-Net architecture employed within these models. This shortcoming often results in overly smooth images and suboptimal imaging quality, indicating areas for future enhancement and research.

\section{Methodology}

\subsection{Multi-Free U-KAN}
Recently, Liu \textit{et al.}~\cite{c1liu2024kan} introduced KANs as a novel alternative to MLPs. While MLPs are effective in modeling complex function mappings and addressing various problems through their multilayer, nonlinear transformations, they have inherent limitations, especially in MRI reconstruction tasks. In such tasks, high-quality images are often recovered from limited and potentially noisy k-space data. MLPs face challenges such as high computational complexity and limited interpretability when handling high-dimensional data. These limitations hinder adequate feature learning, preventing MLPs from capturing detailed features, thus impacting the quality of the reconstructed images. KANs are based on the Kolmogorov-Arnold representation theorem~\cite{c2kolmogorov1957representation}. Unlike MLPs, which use fixed activation functions at neurons and then perform summation for nonlinear activation, KANs deploy learnable activation mechanisms on the connection weights, which are edges, followed by summation. This innovative approach enhances the network's learning capabilities and promotes more flexible feature extraction. A k-layer KAN is formed by nesting multiple KAN layers, mathematically expressed as:
\begin{equation}
\label{eq2}
    \textbf{KAN}(\textbf{Z}) = (\boldsymbol{\Phi}_{k-1}\circ\boldsymbol{\Phi}_{k-2}\circ...\circ\boldsymbol{\Phi}_{1}\circ\Phi_{0})\textbf{Z},
\end{equation}
where \(\mathbf{\Phi}_i\) denotes the \(\boldsymbol{i}\)-th layer of the KAN network. Each KAN layer has \(\mathbf{n}_{in}\) dimensional inputs and \(\mathbf{n}_{out}\) dimensional outputs. The mapping \(\boldsymbol{\Phi}\) is represented by the set \(\{\boldsymbol{\phi_{q,p}}\}\), where \(\mathbf{p}\) ranges from 1 to \(\mathbf{n}_{in}\) and \(\mathbf{q}\) ranges from 1 to \(\mathbf{n}_{out}\). Notably, \(\Phi\) incorporates the learnable activation function \(\boldsymbol{\phi}\) for the \(\mathbf{n}_{in} \times \mathbf{n}_{out}\) dimensional transformation.

The computational results of the KAN network from the \( {k} \)-th layer to the \( {(k+1)} \)-th layer can be expressed through a matrix representation as follows:

\begin{equation}
\resizebox{0.44\textwidth}{!}{$
\label{eq3}
    \textbf{Z}_{{k+1}} = \underbrace{\begin{pmatrix}
  \boldsymbol{\phi}_{k,1,1}(\cdot) & \boldsymbol{\phi}_{k,1,2}(\cdot) & \cdots & \boldsymbol{\phi}_{k,1,n_k}(\cdot)\\
  \boldsymbol{\phi}_{k,2,1}(\cdot) & \boldsymbol{\phi}_{k,2,2}(\cdot) & \cdots & \boldsymbol{\phi}_{k,2,n_k}(\cdot)\\
  \vdots & \vdots & & \vdots \\
  \boldsymbol{\phi}_{k,n_{k+1},1}(\cdot) & \boldsymbol{\phi}_{k,n_{k+1},2}(\cdot) & \cdots & \boldsymbol{\phi}_{k,n_{k+1},n_k}(\cdot)
\end{pmatrix}}_{\Phi_k }\textbf{Z}_k.$}
\end{equation}

\begin{figure*}
\centerline{\includegraphics[width=0.9\textwidth]{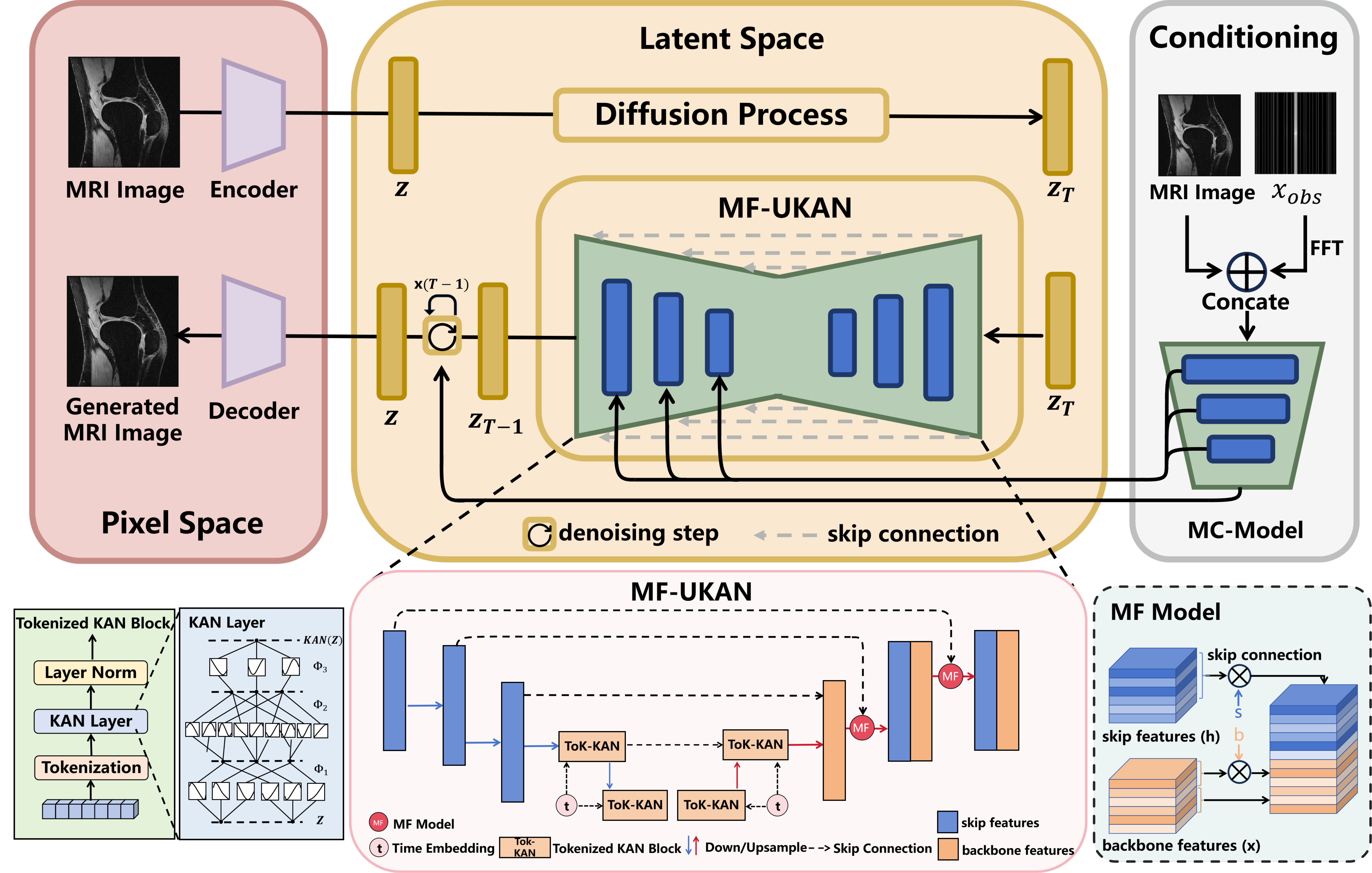}}
\caption{The overall architecture of the TC-KANRecon model. Our model primarily comprises three key components: a VAE module for encoding and decoding images to reduce computational power, a conditional encoder module, MC-Model, for processing conditional images to improve targeted image generation, and a noise prediction backbone, MF-UKAN, which integrates the KAN network and MF model to enhance noise prediction.}
\label{fig1}
\end{figure*}

Inspired by U-KAN~\cite{c3li2024u}, we substitute the U-Net architecture in stable diffusion~\cite{c4rombach2022high} with a U-KAN network. The Tok-KAN module consists of a tokenization layer, a KAN layer, and a Batch Normalization (BN) layer. The structure of the Tok-KAN module is depicted in Fig. \ref{fig1}.

In Tok-KAN, the output feature \(\mathbf{X}_L \in \mathbb{R}^{H_\ell \times W_\ell \times C_\ell}\) from the convolutional stage is reshaped into a flattened 2D sequence of blocks for tokenization as:

\begin{equation}
\label{eq4}
    \mathbf{X}_L^i \in \mathbb{R}^{P^2\cdot C_L}\mid i = 1,...,N,
\end{equation}
where the size of each block is \( P \times P \) and the number of feature blocks is \( N = \frac{(H_L \times W_L)}{P^2}, L = 3 \). The vectorized block \(\mathbf{X}_p\) is then mapped into a potential D-dimensional embedding space using a trainable linear projection \(\mathbf{E} \in \mathbb{R}^{(P^2 \cdot C_L) \times D}\) as:
\begin{equation}
\label{eq5}
    \mathbf{Z}_0 = [\mathbf{X}_L^1\mathbf{E}; 
\mathbf{X}_L^2\mathbf{E};\cdots; 
\mathbf{X}_L^N\mathbf{E}].
\end{equation}
The token sequences are then processed through three KAN layers. Each KAN layer is immediately followed by a BN layer and a ReLU activation function to normalize the data distribution and introduce nonlinear features. Subsequently, Layer Normalization (LN) is applied, and the processed features are passed to the subsequent blocks. The output of the \( k \)-th Tok-KAN block is expressed as:
\begin{equation}
\label{eq6}
    \mathbf{Z}_k = \mathbf{LN(KAN}(\mathbf{Z}_{k-1}))+\mathbf{\mathbf{F}(\mathbf{TE}(t))},
\end{equation}
where \(\mathbf{Z}_k \in \mathbb{R}^{H_k \times W_k \times D_k}\) is the output feature map of the \(k\)-th layer, $\textbf{F}$ is a linear projection, and $\mathbf{{TE(t)}}$ denotes the temporal embedding for a given time step.

The U-KAN, characterized by its U-Net-like structure, facilitates the denoising process through the backbone path and transmits high-frequency detailed features to the decoder part with the help of skip connections. However, this design can overemphasize high-frequency information, potentially weakening the backbone network's ability to capture essential semantic features for denoising. To address this issue, we propose an improved U-KAN module, MF-UKAN, inspired by the MF Model proposed by Zhang \textit{et al.}~\cite{c5zhang2024tc-diffrecon}, which introduces a multi-attention mechanism and two critical scalar modulation factors. The backbone feature scaling factor \( \boldsymbol{b}_l \) aims to enhance the expressive power of the backbone feature map \( \boldsymbol{x}_l \), while the skip feature scaling factor \( \boldsymbol{s}_l \) moderately attenuates the influence of the skip feature map \( \boldsymbol{h}_l \) to avoid excessive high-frequency interference.

For \( \mathbf{b}_l \), we adopt an adaptive adjustment strategy that dynamically adjusts the scaling factor based on statistical information, such as the sample mean:
\begin{small}
\begin{equation}
\label{eq7}
    \boldsymbol{\alpha}_l = (\boldsymbol{b}_l-1)\cdot \frac{\overline{\boldsymbol{x}}_l-\mathbf{Min}(\overline{\boldsymbol{x}}_l)}
{\mathbf{Max}(\overline{\boldsymbol{x}}_l)-\mathbf{Min}(\overline{\boldsymbol{x}}_l)}+1, \\
\overline{\boldsymbol{x}}_l = \frac{1}{\mathbf{C}}\sum_{i=1}^{\mathbf{C}}{\boldsymbol{x}_{l,i}},
\end{equation}
\end{small}
where \( \boldsymbol{x}_{l,i} \) denotes the \(i\)-th channel of the feature mapping \( \boldsymbol{x}_l \), \( \mathbf{C} \) denotes the total number of channels, \( \boldsymbol{\alpha}_l \) is the backbone factor, and \( \boldsymbol{b}_l \) is a scalar constant.

To balance denoising and preserving high-frequency details, we selectively apply the scaling operation to half of the channels of the feature mapping \( \boldsymbol{x}_l \):
\begin{equation}
\label{eq8}
    \boldsymbol{x}_{l,i}^{'}
    \begin{cases}
        \boldsymbol{x}_{l,i}\odot\boldsymbol{\alpha}_l, & \mathrm{if}\ i < \frac{C}{2}, \\
        \boldsymbol{x}_{l,i}, & \mathrm{otherwise.}
    \end{cases}
\end{equation}

For \( \boldsymbol{s}_l \), we employ spectral modulation in the Fourier domain to reduce the low-frequency components of the skip features and thus amplify the high-frequency features:
\begin{equation}
\label{eq9}
    \begin{aligned}
\boldsymbol{\mathcal{F}}(\boldsymbol{h}_{l,i}) &= \boldsymbol{\mathrm{FFT}}(\boldsymbol{h}_{l,i}), \\
\boldsymbol{\mathcal{F}}^{'}(\boldsymbol{h}_{l,i}) &= \boldsymbol{\mathcal{F}}(\boldsymbol{h}_{l,i})\odot \boldsymbol{\beta}_{l,i}, \\
\boldsymbol{h}_{l,i}^{'} &= \boldsymbol{\mathrm{IFFT}}(\boldsymbol{\mathcal{F}}^{'}(\boldsymbol{h}_{l,i})), \\
\boldsymbol{\alpha}_{l,i}(r) &= 
\begin{cases}
\boldsymbol{s}_l, & \text{if } \boldsymbol{r} < \boldsymbol{r}_{thresh}; \\
1, & \text{otherwise,}
\end{cases}
\end{aligned}
\end{equation}
where \textbf{FFT} and \textbf{IFFT} are the Fourier transform and inverse Fourier transform, respectively. Pixel-level multiplication is denoted by \(\odot\), and \(\boldsymbol{\beta}_{l,i}\) is a Fourier mask designed based on the size of the Fourier coefficients, implementing the frequency-dependent scale factor \( \boldsymbol{s}_l \). The radius is represented as \( \boldsymbol{r} \), while \( \boldsymbol{r}_{thresh} \) indicates the threshold frequency. Finally, we combine the enhanced skip feature map with the finely tuned backbone feature map to serve as inputs for subsequent layers in the U-Net architecture.
\subsection{MC-Model}
To enhance the directional generation of MR images, we use fully sampled k-space information as a conditional guide. Initially, we apply an inverse Fourier transform to the fully sampled k-space data, denoted as \(\boldsymbol{x}_{obs}\), and subsequently fuse it with the MR images through concatenation. This process is represented by Eq. \ref{eq10}:
\begin{equation}
\label{eq10}
        \mathbf{\tilde{X}} = \boldsymbol{\mathrm{IFFT}}( \boldsymbol{x}_{obs}) \boldsymbol{\oplus} \mathbf{X},
\end{equation}
where \textbf{IFFT} denotes the inverse Fourier transform, \(\mathbf{\oplus}\) signifies concatenation, and \(\mathbf{\tilde{X}}\) represents the conditional information post-concatenation. Noise is then added to \(\mathbf{\tilde{X}}\) to obtain \(\mathbf{\tilde{X}}'\).

To maintain consistency with the diffusion backbone's input, which is potential space, we compress \(\mathbf{\tilde{X}}'\) using an additional convolution module. This module employs a kernel size of 3, a step size of 1, and a padding of 1. The compression process involves an initial convolutional layer, followed by a ReLU activation function and a group normalization layer. This sequence is executed three times to produce a feature map with the same shape as the diffusion backbone, which is then passed to subsequent modules. The final feature vector \(\mathbf{Z}\) is described by Eq. \ref{eq11}:
\begin{equation}
\label{eq11}
    \mathbf{Z} = \mathbf{Conv}(\mathbf{\tilde{X^{}}}{'}),
\end{equation}
where \(\text{\textbf{Conv}}(\cdot)\) denotes the convolution operation. Unlike alternative approaches that utilize cross-attention, tandem, or CLIP image encoders to extract and transmit high-level semantic information from images, we employ the encoder stage of MF-UKAN. This stage consists of three downsampling blocks and two Tok-KAN modules, as illustrated in Fig. \ref{fig1}. Each convolution block comprises a convolutional layer, a batch normalization layer, and a ReLU activation function, with a kernel size of 3, a step size of 1, and a padding of 1.

This module processes the fully sampled k-space information and the Fourier-transformed MR images information. It then transmits the resulting latent representations through the downsampled blocks and intermediate blocks, meticulously recording all intermediate feature mappings. Finally, these recorded feature mappings are introduced into the upsampled segment of the diffusion backbone.

\subsection{Dynamic Clipping Strategy}
The clipping threshold is crucial in balancing image quality and diversity during the generation process. A higher clipping threshold retains more predictive noise and variations, resulting in diverse styles and details in the generated images, but it may also introduce unwanted noise or blurriness, affecting the overall image quality. Conversely, a lower clipping threshold reduces noise, producing clearer and more stable images, but excessive restriction can lead to a lack of detail and diversity, making the images appear overly smooth. Therefore, selecting an appropriate clipping threshold is essential to achieve an optimal balance between image quality and diversity.

\begin{figure*}
\centerline{\includegraphics[width=1.0\textwidth]{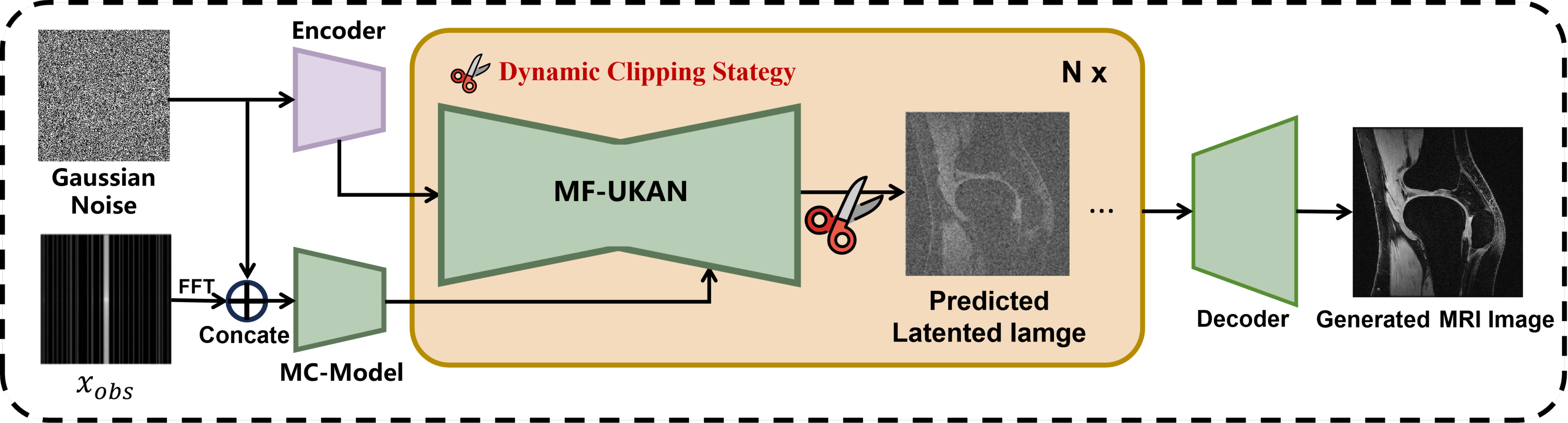}}
\caption{Details of the dynamic cropping strategy. After completing model training, a dynamically adaptive, linearly decreasing cropping technique that adapts dynamically at each step of the sampling process. The adjustment is based on the current progress of each sampling step. The adjustment is based on the current progress of each sampling step. It is applied to the noise estimates generated by the MF-UKAN module, resulting in MR images with improved visual effects.}
\label{fig2}
\end{figure*}

Recently, Sun \textit{et al.}~\cite{sun2024bsldmeffectivebonesuppression} proposed an innovative temporal adaptive thresholding strategy to replace the conventional fixed clipping approach, which strictly confines the predicted variable \(\mathbf{x}\) within the interval \([-1,1]\). This new strategy involves dynamically adjusting the clipping interval boundaries based on the current sampling step \(\mathbf{t}\), defining a varying interval \([\mathbf{-s, s}]\), where \(\mathbf{s}\) evolves with the sampling process. They formulated a linear model as shown in Eq. \ref{eq12}:
\begin{equation}
\label{eq12}
    \mathbf{{s} = \omega \cdot {t}+b}.
\end{equation}
This strategy allows for flexible adjustment of the clipping threshold according to the model's training progress and the stage of image generation. It provides greater freedom during the initial stages of image structure formation, resulting in richer visual features. However, during the early stages of diffusion model sampling, when the model generates blurry, low-resolution initial image structures from pure noise, a larger clipping threshold is needed to increase freedom and reduce information loss, aiding in capturing overall shapes. As sampling progresses and image resolution and details increase, the clipping threshold should gradually decrease to limit noise and enhance image finesse and quality. Consequently, we have refined the linear model to Eq. \ref{eq13}:
\begin{equation}
\label{eq13}
    \mathbf{{s} = - \omega \cdot {t}+b}.
\end{equation}
This revised model allows the clipping threshold \(\mathbf{s}\) to linearly decrease as the sampling step \(\mathbf{t}\) increases. This dynamic clipping strategy preserves image diversity during the initial stages of diffusion model sampling and retains image details and precision as sampling deepens. Consequently, it produces images that better align with real-world counterparts, as illustrated in Fig. \ref{fig2}

\section{Experiment}

\subsection{Datasets}
The performance of the TC-KANRecon algorithm was evaluated using the fastMRI~\cite{d1zbontar2018fast} and SKM-TEA~\cite{d2desai2022skm} datasets. The fastMRI dataset includes raw k-space data and DICOM images of knee, brain, prostate, and chest scans, accompanied by masked test sets. This dataset is specifically designed to facilitate MRI reconstruction and aims to accelerate the generation of MR images using artificial intelligence techniques. For this study, we generated the undersampled mask \(\boldsymbol{\mathcal{M}}\) using the mask function provided in the fastMRI challenge. We concentrated on the single-coil knee dataset, which contains data from 1,172 subjects, each with approximately 35 slices. From this dataset, we used data from 973 subjects, about 34,055 slices, for training and data from 199 subjects, about 6,965 slices, for evaluation.

The SKM-TEA dataset comprises DICOM images from 155 patients scanned using two Tesla 3T GE MR750 scanners. This dataset pairs raw quantitative knee MRI data with dense labeling of the original tissues and cases, allowing for comprehensive analysis. For our experiments, we selected single-coil data from 515 subjects, each with approximately 160 slices. Of these, data from 129 subjects, about 20,640 slices, were used for training, while data from 26 subjects, about 4,160 slices, were used for evaluation.

\subsection{Experimental Details}
The TC-KANRecon method is implemented using PyTorch 2.0 with CUDA 11.8 and trained on two NVIDIA A800 GPUs. Prior to training, we centrally cropped the fastMRI and SKM-TEA datasets to 256$\times$256. The training process includes a VAE component and a diffusion component. The VAE component is trained with a batch size of 8 for 48 hours. Conversely, the diffusion component is trained with a batch size of 32 for 36 hours. Both components are optimized using the Adam optimizer. The learning rate is set to 0.0001 for the VAE component and 0.000025 for the diffusion component.

\subsection{Evaluation Metrics}
To thoroughly assess the performance of our model, we employed three evaluation metrics: Peak Signal-to-Noise Ratio (PSNR), Structural Similarity Index (SSIM), and Normalized Mean Square Error (NMSE). PSNR measures the peak signal to noise ratio, indicating image quality by comparing signal energy to noise. 
SSIM compares image similarity in terms of brightness, contrast, and structure, with higher values meaning the reconstructed image looks more like the original. 
NMSE normalizes the mean square error against the original image's error, with lower values indicating higher quality. 
These metrics provide a comprehensive assessment of our model's performance in reconstructing MRI images accurately.

\subsection{Comparison Experiments}
\begin{figure*}
\centerline{\includegraphics[width=1.0\textwidth]{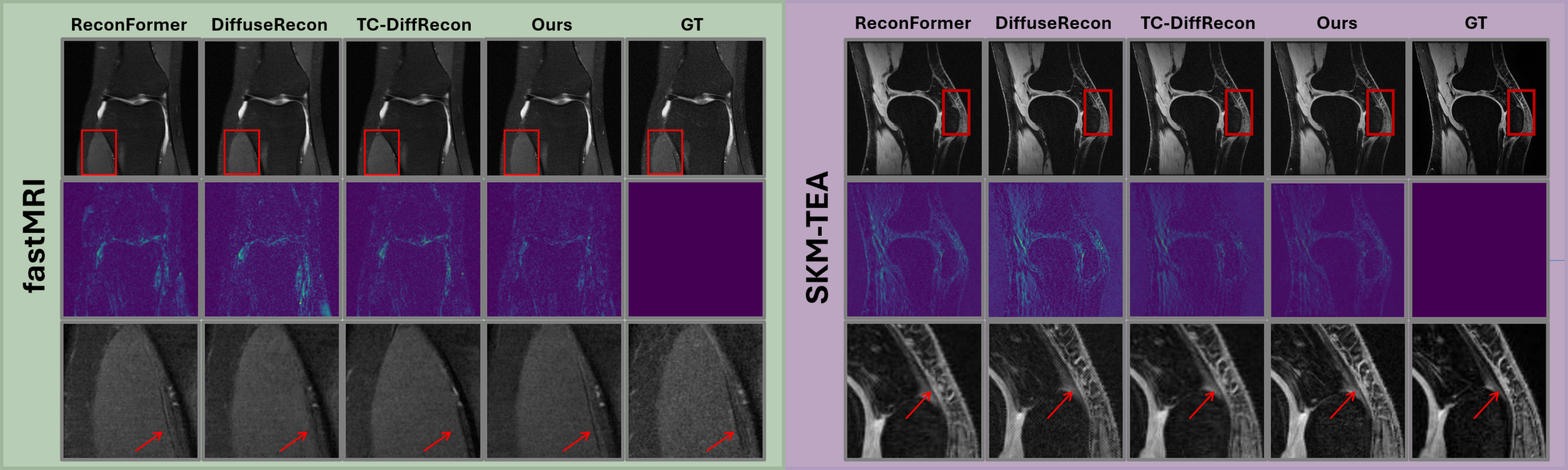}}
\caption{Compared with advanced reconstruction models on the fastMRI and SKM-TEA datasets in terms of visual effects. Enlarged detail images reveal that our approach better preserves texture details, indicating a stronger ability to restore fine features.}
\label{fig3}
\end{figure*}

As shown in Tables \ref{tab1} and \ref{tab2}, we conducted comparative experiments with Zero Filled (ZF) and other state-of-the-art models for MRI reconstruction on the fastMRI and SKM-TEA datasets to validate the efficacy of the TC-KANRecon model. The experimental results indicate that the TC-KANRecon model significantly outperforms its counterparts in terms of PSNR, SSIM, and NMSE metrics. The innovative network structure and diversified feature processing strategies of the TC-KANRecon model enhance its adaptability and robustness in MRI reconstruction tasks. Consequently, it surpasses existing models in all evaluated metrics, demonstrating its superior performance. The comparative results with other state-of-the-art models, specifically for an Acceleration Factor (AF) of 4, are illustrated in Fig. \ref{fig3}. Given the consistent results across different single-coil datasets and various acceleration factors, we focus our detailed analysis on the reconstruction performance of the model using the fastMRI dataset with AF = 4, both in comparison and ablation experiments.

Compared to the KIKI-net model~\cite{eo2018kiki}, which struggles with high-frequency detail processing and denoising due to its lack of effective strategies for capturing fine details and suppressing noise, the TC-KANRecon model demonstrates superior performance. The innovative texture coordination strategy and multi-attention mechanism of TC-KANRecon enhance detail restoration helping the model excel in restoring high-frequency details and effectively managing noise, significantly improving overall image quality. Specifically, when compared with KIKI-net, TC-KANRecon increases the PSNR from 27.01 to 30.50 and the SSIM from 0.613 to 0.817. In terms of NMSE metrics, KIKI-net scores 0.0354, while TC-DiffRecon achieves a lower value of 0.0211.

Compared to the D5C5 model~\cite{b9schlemper2017deep}, which shows some advantages in MRI reconstruction, D5C5's static feature extraction methods result in partial detail loss, affecting detail restoration and overall image quality. The PSNR for D5C5 is 27.74, while TC-KANRecon achieves 30.50. The SSIM for D5C5 is 0.632, lower than TC-KANRecon's 0.817. NMSE metrics for U-Net are 0.0289, while TC-KANRecon scores 0.0211. TC-KANRecon, with its multi-attention mechanism and dynamic clipping strategy, significantly outperforms D5C5 in reducing details and overall image quality, demonstrating superior image detail restoration capability.

Compared to the U-Net model~\cite{b5ronneberger2015u}, although U-Net performs well in many image processing tasks, it struggles with capturing high-frequency details and dealing with noise in MR images due to its encoder-decoder structure, which can result in the loss of detail during feature extraction. For instance, the PSNR of the U-Net model is 27.68, while the TC-KANRecon model achieves 30.50. Similarly, the SSIM for U-Net is 0.631, significantly lower than TC-KANRecon's 0.817. In terms of NMSE, U-Net scores 0.0295, whereas TC-KANRecon registers a lower 0.0211. The TC-KANRecon model, incorporating the MF Model, optimizes the balance of skip connections and main features, substantially enhancing reconstruction accuracy.

Compared to the OUCR model~\cite{b14guo2021over}, TC-KANRecon demonstrates significant advantages in MRI reconstruction. While the OUCR model captures both local and global features by combining overcomplete and undercomplete branches, it falls short in detail restoration and overall image quality. Specifically, the OUCR model achieves a PSNR of 28.08, whereas TC-KANRecon attains 30.05. The SSIM for the OUCR model is 0.633, markedly lower than TC-KANRecon's 0.817. Additionally, the NMSE metric for the OUCR model is 0.0277, compared to TC-KANRecon's improved performance of 0.0211. TC-KANRecon leverages an innovative conditional guided diffusion model, which integrates the MF-UKAN module and a dynamic clipping strategy, effectively balancing image denoising and structure preservation.

\begin{table*}
\centering
\footnotesize 
\caption{Comparison of TC-KANRecon with other state-of-the-art models on the public fastMRI dataset.}
\label{tab1}
\setlength{\tabcolsep}{3pt} 
\begin{tabularx}{\textwidth}{c|>{\centering\arraybackslash}X>{\centering\arraybackslash}X>{\centering\arraybackslash}X>{\centering\arraybackslash}X|>{\centering\arraybackslash}X>{\centering\arraybackslash}X>{\centering\arraybackslash}X>{\centering\arraybackslash}X|>{\centering\arraybackslash}X >{\centering\arraybackslash}X >{\centering\arraybackslash}X c}
\toprule
\multicolumn{13}{c}{fastMRI~\cite{d1zbontar2018fast}}                                                      \\ 
\hline
Method                                                                       & \multicolumn{4}{c|}{PSNR($\uparrow$)}                                     & \multicolumn{4}{c|}{SSIM($\uparrow$)}                                     & \multicolumn{4}{c}{NMSE($\downarrow$)}                                         \\ 
\hline
AF Factor                                                                     & 4             & 6              & 8              & 10             & 4             & 6              & 8              & 10           & 4             & 6              & 8              & 10               \\ 
\hline
ZF                                                                           & 22.91          & 20.79          & 18.37          & 15.07          & 0.549          & 0.482          & 0.414          & 0.349          & 0.0436          & 0.0488          & 0.0567          & 0.0719           \\
KIKI-net~\cite{eo2018kiki}                                                                     & 27.01             & 26.55             & 25.08            & 24.06             & 0.613              & 0.503              & 0.492              & 0.412              & 0.0354              & 0.0397               & 0.0462               & 0.0542                \\
D5C5~\cite{b9schlemper2017deep}                                                                         & 27.74              & 26.56              & 26.12              & 24.60             & 0.632              & 0.529              & 0.512              & 0.445              & 0.0289               & 0.0374               & 0.0440               & 0.0532                \\
U-Net~\cite{b5ronneberger2015u}                                                                         & 27.68          & 26.23          & 25.83          & 24.49          & 0.631          & 0.520          & 0.511          & 0.436          & 0.0295          & 0.0382          & 0.0430          & 0.0520           \\
OUCR~\cite{b14guo2021over}                                                                         & 28.08          & 27.66          & 26.23          & 25.31          & 0.633          & 0.534          & 0.516          & 0.482          & 0.0277          & 0.0368          & 0.0389          & 0.0486           \\
DiffuseRecon~\cite{b24peng2022towards}                                                                 & 28.27          & 27.83          & 26.60          & 25.90          & 0.635          & 0.586          & 0.549          & 0.528          & 0.0272          & 0.0360          & 0.0381          & 0.0466           \\
ReconFormer~\cite{guo2023reconformer}                                                                  & 28.62          & 28.01          & 26.64          & 26.08          & 0.643          & 0.590          & 0.529          & 0.519          & 0.0263          & 0.0348          & 0.0374          & 0.0480           \\
TC-DiffRecon~\cite{c5zhang2024tc-diffrecon}                                                                 & 29.78              & 28.77              & 27.35              & 26.49              & 0.742              & 0.637              & 0.591              & 0.562              & 0.0247              & 0.0313               & 0.0338               & 0.0415                \\
\textbf{TC-KANRecon (Ours)} & \textbf{30.50} & \textbf{29.16} & \textbf{27.93} & \textbf{26.83} & \textbf{0.817} & \textbf{0.691} & \textbf{0.624} & \textbf{0.593} & \textbf{0.0211} & \textbf{0.0273} & \textbf{0.0309} & \textbf{0.0371}  \\
\bottomrule 
\end{tabularx}
\end{table*}

\begin{table*}
\centering
\footnotesize 
\caption{Comparison of TC-KANRecon with other state-of-the-art models on the public SKM-TEA dataset.}
\label{tab2}
\setlength{\tabcolsep}{3pt} 
\begin{tabularx}{\textwidth}{c|>{\centering\arraybackslash}X>{\centering\arraybackslash}X>{\centering\arraybackslash}X>{\centering\arraybackslash}X|>{\centering\arraybackslash}X>{\centering\arraybackslash}X>{\centering\arraybackslash}X>{\centering\arraybackslash}X|>{\centering\arraybackslash}X >{\centering\arraybackslash}X >{\centering\arraybackslash}X c}
\toprule
\multicolumn{13}{c}{SKM-TEA~\cite{d2desai2022skm}}                                                      \\ 
\hline
Method                                                                       & \multicolumn{4}{c|}{PSNR($\uparrow$)}                                     & \multicolumn{4}{c|}{SSIM($\uparrow$)}                                     & \multicolumn{4}{c}{NMSE($\downarrow$)}                                         \\ 
\hline
AF Factor                                                                    & 4              & 6              & 8              & 10             & 4              & 6              & 8              & 10             & 4               & 6               & 8               & 10               \\ 
\hline
ZF                                                                           & 23.63          & 22.09          & 20.06          & 16.72          & 0.653          & 0.587          & 0.529          & 0.455          & 0.0361          & 0.0422          & 0.0511          & 0.0641           \\
KIKI-net~\cite{eo2018kiki}                                                  & 28.16          & 26.96          & 25.80          & 25.11          & 0.716          & 0.620          & 0.601          & 0.543          & 0.0286          & 0.0350          & 0.0434          & 0.0510           \\
D5C5~\cite{b9schlemper2017deep}                                             & 28.87          & 28.24          & 27.35          & 27.01          & 0.731          & 0.654          & 0.637          & 0.586          & 0.0221          & 0.0260          & 0.0384          & 0.0460           \\
U-Net~\cite{b5ronneberger2015u}                                             & 28.52          & 27.79          & 26.73          & 26.34          & 0.723          & 0.645          & 0.631          & 0.571          & 0.0254          & 0.0336          & 0.0410          & 0.0493           \\
OUCR~\cite{b14guo2021over}                                                  & 29.26          & 28.90          & 28.06          & 27.24          & 0.745          & 0.656          & 0.642          & 0.603          & 0.0165          & 0.0247          & 0.0320          & 0.0387           \\
DiffuseRecon~\cite{b24peng2022towards}                                      & 29.51          & 29.13          & 28.20          & 27.49          & 0.749          & 0.712          & 0.682          & 0.668          & 0.0152          & 0.0232          & 0.0303          & 0.0352           \\
ReconFormer~\cite{guo2023reconformer}                                       & 29.74          & 29.43          & 28.77          & 28.37          & 0.765          & 0.724          & 0.693          & 0.654          & 0.0149          & 0.0216          & 0.0278          & 0.0326           \\
TC-DiffRecon~\cite{c5zhang2024tc-diffrecon}                                 & 30.59          & 30.26          & 29.50          & 29.22          & 0.852          & 0.771          & 0.732          & 0.708          & 0.0136          & 0.0198          & 0.0254          & 0.0311           \\
\textbf{TC-KANRecon (Ours)}                                                   & \textbf{31.48} & \textbf{30.93} & \textbf{30.12} & \textbf{29.74} & \textbf{0.893} & \textbf{0.822} & \textbf{0.785} & \textbf{0.741} & \textbf{0.0118} & \textbf{0.0189} & \textbf{0.0242} & \textbf{0.0274}  \\
\bottomrule
\end{tabularx}
\end{table*}
Compared to the ReconFormer model~\cite{guo2023reconformer}, which excels in capturing long-range dependencies using a Transformer structure, TC-KANRecon demonstrates improved performance in handling complex details of MR images. While ReconFormer effectively captures global information, it often overlooks local detail features, leading to suboptimal performance. In contrast, TC-KANRecon combines the MF Model and the Tok-KAN module to enhance feature extraction and model interpretability. This results in superior performance in metrics such as PSNR and SSIM. Specifically, TC-KANRecon achieves a PSNR of 30.50, compared to ReconFormer's 28.62. Additionally, the SSIM for TC-KANRecon reaches 0.817, while ReconFormer attains 0.643. Regarding NMSE metrics, ReconFormer scores 0.0263, whereas TC-KANRecon achieves a lower value of 0.0211.

Compared to the DiffuseRecon model~\cite{b24peng2022towards}, which achieves better reconstruction through the diffusion model, TC-KANRecon proves superior in handling complex textures. DiffuseRecon is prone to artifacts that lead to texture inconsistency and detail loss. For instance, the NMSE of DiffuseRecon is 0.0272, whereas TC-KANRecon achieves a lower NMSE of 0.0211, indicating higher image quality. The TC-KANRecon model combines a multi-attention mechanism with adaptive feature adjustment strategies, enhancing feature extraction and interpretability. This allows it to excel in capturing complex pathological features and restoring details. In contrast, the TC-DiffRecon model, although it addresses issues like image fragmentation caused by over-smoothing and improves overall image quality and consistency, still struggles with high-frequency details and complex textures. TC-KANRecon significantly improves the ability to capture complex pathological features and overall image reconstruction quality through its innovative Tok-KAN module and dynamic clipping strategy. Specifically, TC-KANRecon improves PSNR by 0.72 and SSIM by 0.076 on the STM-TEA dataset. DiffuseRecon's NMSE is 0.0272, while TC-DiffRecon's is 0.0247. In summary, TC-KANRecon excels in reproducing image details and handling complex textures, demonstrating higher robustness and stability. It maintains excellent performance across different acceleration factors, outperforming other models in various key metrics.

\subsection{Validation of Model Generalizability}
To further validate the broad applicability of our proposed TC-KANRecon model, we conducted a detailed comparison of its generalizability against several typical reconstruction models using two public datasets: fastMRI and SKM-TEA. As illustrated in Table \ref{tab3}, we trained our model with AF of 6× and 8×, and used undersampled images with acceleration factors of 10× and 4× as inputs. The results demonstrate that, in contrast to models optimized for a specific AF, TC-KANRecon shows significant performance advantages, thereby proving its robust versatility. Furthermore, when compared with diffusion-based models known for their extensive applicability, our model achieved the highest quality reconstruction outcomes. This underscores TC-KANRecon's superior performance in handling a range of undersampling scenarios.

\subsection{Ablation Study}
We conducted ablation experiments to evaluate the impact of the MF Model, Tok-KAN module, and dynamic clipping strategy on the TC-KANRecon model's performance using the fastMRI and SKM-TEA datasets, as shown in Tables \ref{tab4} and \ref{tab5}. We compared the performance metrics, PSNR, SSIM, and NMSE, of the complete TC-KANRecon model with versions from which each component was individually removed. The results reveal that removing any of these modules significantly degrades the model's reconstruction performance, underscoring their essential roles in the overall effectiveness of our model. Consequently, the TC-KANRecon model demonstrates enhanced adaptability and robustness in MRI reconstruction tasks, attributed to its innovative network structures and diverse feature processing strategies.

\subsubsection{Impact after removing the MF Model}
The MF Model significantly enhances the model's ability to extract features and reduce noise by utilizing a multi-head attention mechanism and feature scaling strategy. When this module is removed, the model's performance deteriorates markedly in both high-frequency detail restoration and noise reduction. Specifically, this decline is evident in the substantial decrease in PSNR values from 30.50 to 28.67 and SSIM values from 0.817 to 0.674. Additionally, NMSE increases from 0.0211 to 0.0287. All these results underscore the critical role of the MF Model for the overall model in optimizing feature representation and the denoising effect.

\begin{table*}
\centering
\caption{Using 10$\times$ and 4$\times$ undersampling for training on the fastMRI and SKM-TEA datasets, while the 6$\times$ and 8$\times$ for sampling.}
\label{tab3}
\begin{tabularx}{1\textwidth}{c|>{\centering}X>{\centering}X|>{\centering}X>{\centering}X|>{\centering}X>{\centering}X|>{\centering}X>{\centering}X|>{\centering}X>{\centering}X| >{\centering}Xc} 
\toprule
$~$     &
\multicolumn{6}{c|}{{fastMRI~\cite{d1zbontar2018fast}}}     &
\multicolumn{6}{c}{{SKM-TEA~\cite{d2desai2022skm}}}                                                                                                                                                                                                                                                                  \\ 
\hline
Method                                                                       & \multicolumn{2}{c|}{PSNR($\uparrow$)}                                     & \multicolumn{2}{c|}{SSIM($\uparrow$)}                                     & \multicolumn{2}{c|}{NMSE($\downarrow$)}         &
\multicolumn{2}{c|}{PSNR($\uparrow$)}                                     & \multicolumn{2}{c|}{SSIM($\uparrow$)}                                     & \multicolumn{2}{c}{NMSE($\downarrow$)}                      \\ 
\hline
AF Factor                                                                           & \makebox[0.05\textwidth][c]{8$\times$$\rightarrow$4$\times$}            & \makebox[0.05\textwidth][c]{6$\times$$\rightarrow$10$\times$}
& \makebox[0.05\textwidth][c]{8$\times$$\rightarrow$4$\times$}            & \makebox[0.05\textwidth][c]{6$\times$$\rightarrow$10$\times$}
& \makebox[0.05\textwidth][c]{8$\times$$\rightarrow$4$\times$}            & \makebox[0.05\textwidth][c]{6$\times$$\rightarrow$10$\times$}
& \makebox[0.05\textwidth][c]{8$\times$$\rightarrow$4$\times$}            & \makebox[0.05\textwidth][c]{6$\times$$\rightarrow$10$\times$}
& \makebox[0.05\textwidth][c]{8$\times$$\rightarrow$4$\times$}            & \makebox[0.05\textwidth][c]{6$\times$$\rightarrow$10$\times$}
& \makebox[0.05\textwidth][c]{8$\times$$\rightarrow$4$\times$}            & \makebox[0.05\textwidth][c]{6$\times$$\rightarrow$10$\times$}

\\ 
\hline
KIKI-net~\cite{eo2018kiki}                                                                     & 26.38               & 22.06                 & 0.522               & 0.393                 & 0.0383              & 0.0561               & 27.04               & 22.96                 & 0.614               & 0.512                 & 0.0348              & 0.0521               \\
D5C5~\cite{b9schlemper2017deep}                                                                         & 26.64               & 23.60                 & 0.534               & 0.429                 & 0.0361              & 0.0554               & 27.87               & 24.24                 & 0.653               & 0.557                 & 0.0314              & 0.0492               \\
U-Net~\cite{b5ronneberger2015u}                                                                       & 26.53               & 23.49                 & 0.528               & 0.385                 & 0.0377              & 0.0542               & 27.62               & 23.80                 & 0.645               & 0.530                 & 0.0335              & 0.0487               \\
OUCR~\cite{b14guo2021over}                                                                         & 27.93               & 24.31                 & 0.564               & 0.478                 & 0.0354              & 0.0501               & 28.80               & 25.04                 & 0.672               & 0.582                 & 0.0282              & 0.0421               \\
ReconFormer~\cite{guo2023reconformer}                                                                 & 28.20               & 25.21                 & 0.596               & 0.490                 & 0.0347              & 0.0492               & 29.47               & 25.73                 & 0.711               & 0.630                 & 0.0258              & 0.0395               \\
\begin{tabular}[c]{@{}c@{}}\scriptsize\textbf{TC-KANRecon (Ours)} \end{tabular}& \textbf{30.50}      & \textbf{26.83}        & \textbf{0.817}      & \textbf{0.593}        & \textbf{0.0211}     & \textbf{0.0371}      & \textbf{31.48}      & \textbf{29.74}        & \textbf{0.893}      & \textbf{0.741}        & \textbf{0.0118}     & \textbf{0.0274}    \\
\bottomrule
\end{tabularx}
\end{table*}

\begin{table*}
\centering
\caption{TC-KANRecon ablation experiments on the public fastMRI dataset.}
\label{tab4}
\begin{tabularx}{\textwidth}{c|>{\centering}X>{\centering}X>{\centering}X>{\centering}X|>{\centering}X>{\centering}X>{\centering}X>{\centering}X|>{\centering}X >{\centering}X >{\centering}X c} 
\toprule
\multicolumn{13}{c}{{fastMRI~\cite{d1zbontar2018fast}}}                                                                                                                                                                                                                                                                  \\ 
\hline
Method                                                                       & \multicolumn{4}{c|}{PSNR($\uparrow$)}                             & \multicolumn{4}{c|}{SSIM($\uparrow$)}                             & \multicolumn{4}{c}{NMSE($\downarrow$)}                                 \\ 
\hline
AF Factor                                                                    & 4              & 6              & 8              & 10             & 4              & 6              & 8              & 10             & 4               & 6               & 8               & 10               \\ 
\hline
w/o MF Model                                                                         & 28.67          & 28.21          & 26.59          & 25.97          & 0.674          & 0.621          & 0.597          & 0.559          & 0.0287          & 0.0364          & 0.0454          & 0.0506           \\
w/o Tok-KAN module                                                                 & 28.34          & 28.02          & 26.24          & 25.70          & 0.642          & 0.589          & 0.571          & 0.520          & 0.0326          & 0.0398          & 0.0497          & 0.0574           \\
w/o Dynamic Clipping Strategy                                                                  & 29.86          & 28.89          & 27.66          & 26.64          & 0.754          & 0.642          & 0.598          & 0.571          & 0.0266          & 0.0324          & 0.0343          & 0.0431           \\
\begin{tabular}[c]{@{}c@{}}\textbf{TC-KANRecon (Ours)}\end{tabular} & \textbf{30.50} & \textbf{29.16} & \textbf{27.93} & \textbf{26.83} & \textbf{0.817} & \textbf{0.691} & \textbf{0.624} & \textbf{0.593} & \textbf{0.0211} & \textbf{0.0273} & \textbf{0.0309} & \textbf{0.0371}  \\
\bottomrule
\end{tabularx}
\end{table*}

\begin{table*}
\centering
\caption{TC-KANRecon ablation experiments on the public SKM-TEA dataset.}
\label{tab5}
\begin{tabularx}{\textwidth}{c|>{\centering}X>{\centering}X>{\centering}X>{\centering}X|>{\centering}X>{\centering}X>{\centering}X>{\centering}X|>{\centering}X >{\centering}X >{\centering}X c} 
\toprule
\multicolumn{13}{c}{{SKM-TEA~\cite{d2desai2022skm}}}                                                                                                                                                                                                                                                                  \\ 
\hline
Method                                                                       & \multicolumn{4}{c|}{PSNR($\uparrow$)}                             & \multicolumn{4}{c|}{SSIM($\uparrow$)}                             & \multicolumn{4}{c}{NMSE($\downarrow$)}                                 \\ 
\hline
AF Factor                                                                    & 4              & 6              & 8              & 10             & 4              & 6              & 8              & 10             & 4               & 6               & 8               & 10               \\ 
\hline
w/o MF Model                                                                         & 29.12          & 28.77          & 28.26          & 27.74          & 0.811          & 0.724          & 0.674          & 0.620          & 0.0179          & 0.0230          & 0.0304          & 0.0331           \\
w/o Tok-KAN module                                                                 & 28.78          & 28.45          & 27.84          & 27.31          & 0.765          & 0.671          & 0.646          & 0.589          & 0.0217          & 0.0256          & 0.0323          & 0.0376           \\
w/o Dynamic Clipping Strategy                                                                  & 30.18          & 29.24          & 28.81          & 28.40          & 0.845          & 0.768          & 0.724          & 0.675          & 0.0144          & 0.0201          & 0.0284          & 0.0298           \\
\begin{tabular}[c]{@{}c@{}}\textbf{TC-KANRecon (Ours)}\end{tabular} & \textbf{31.48} & \textbf{30.93} & \textbf{30.12} & \textbf{29.74} & \textbf{0.893} & \textbf{0.822} & \textbf{0.785} & \textbf{0.741} & \textbf{0.0118} & \textbf{0.0189} & \textbf{0.0242} & \textbf{0.0274}  \\
\bottomrule
\end{tabularx}
\end{table*}

\subsubsection{Impact after removing the Tok-KAN Module}
The Tok-KAN module enhances the flexibility of feature extraction and representation by applying the Kolmogorov-Arnold expression theorem, which significantly boosts the model's image reconstruction quality. When this module is removed, the model's capacity to capture complex pathological features and detailed information is diminished. This reduction is reflected in the significant drop in SSIM values from 0.817 to 0.642 and the increase in NMSE from 0.0211 to 0.0326. These changes highlight the essential role of the Tok-KAN module in improving feature extraction flexibility and overall image reconstruction quality.

\subsubsection{Impact after removing the Dynamic Clipping Strategy}
The dynamic clipping strategy enhances image diversity and quality while maintaining numerical stability by adjusting the boundaries of the cropping interval during the sampling step. When this strategy is removed, the model exhibits erratic performance in handling various sampling steps, resulting in reduced image reconstruction quality. This decline is evidenced by a decrease in PSNR from 30.50 to 29.86, a drop in SSIM from 0.817 to 0.754, and an increase in NMSE from 0.0211 to 0.0266. These results underscore the critical importance of the dynamic clipping strategy in improving image quality and maintaining the model's stability.

\section{Conclusion}
This paper introduces an innovative deep learning framework, TC-KANRecon, designed to accelerate and enhance the quality of MRI image reconstruction. By integrating the MF-UKAN module and a dynamic clipping strategy, the framework effectively balances denoising and structural preservation, significantly improving the model's robustness and retention of structural details in complex noise environments. Additionally, the MC-Model module incorporates fully sampled k-space information, further enhancing the realism and richness of reconstructed images. Experimental results on two large-scale knee MRI datasets, fastMRI and SKM-TEA, demonstrate that TC-KANRecon outperforms existing methods in terms of image quality, achieving an optimal balance between denoising and detail preservation, and showcasing strong generalization capabilities. The significance of this work lies in providing a highly efficient and accurate MRI image reconstruction method for clinical diagnosis, potentially optimizing the use of medical resources and enhancing diagnostic accuracy.

Despite these significant advancements, there are still some limitations and areas for improvement. One major challenge is the model’s reliance on large-scale datasets for training, which can be problematic in situations where such extensive data is unavailable. Future work will explore techniques such as data augmentation, transfer learning, and synthetic data generation to mitigate this issue. Another important consideration is the computational efficiency of the TC-KANRecon model. As deep learning models grow increasingly complex, it becomes crucial to ensure they can be deployed in real-time clinical environments without significant delays. Future research will focus on optimizing the model architecture to accelerate inference time while maintaining high quality.


\end{document}